\documentclass[aps,prb,twocolumn,superscriptaddress]{revtex4-1}
\usepackage{graphicx}
\usepackage{color}
\usepackage{float}
\begin{document}

\title{Tailored nanoscale plasmon-enhanced vibrational electron spectroscopy}

\author{Luiz~H.~G.~Tizei}
\email{luiz.galvao-tizei@u-psud.fr}
\affiliation{Laboratoire de Physique des Solides, Universit\'e Paris-Sud, CNRS-UMR 8502, Orsay 91405, France}

\author{Vahagn~Mkhitaryan}
\affiliation{ICFO-Institut de Ciencies Fotoniques, The Barcelona Institute of Science and Technology, 08860 Castelldefels (Barcelona), Spain}

\author{Hugo~Louren\c{c}o-Martins}
\affiliation{Laboratoire de Physique des Solides, Universit\'e Paris-Sud, CNRS-UMR 8502, Orsay 91405, France}

\author{Leonardo~Scarabelli}
\affiliation{CIC biomaGUNE and Ciber-BBN, Paseo de Miram\'on 182, 20014 Donostia-San Sebasti\'an, Spain}
\affiliation{Department of Chemistry and Biochemistry, University of California, Los Angeles, Los Angeles, California 90095, USA}

\author{Kenji~Watanabe}
\affiliation{National Institute for Materials Science, Namiki 1-1, Tsukuba, Ibaraki 305-0044, Japan}

\author{Takashi~Taniguchi}
\affiliation{National Institute for Materials Science, Namiki 1-1, Tsukuba, Ibaraki 305-0044, Japan}

\author{Marcel~Tenc\'e}
\affiliation{Laboratoire de Physique des Solides, Universit\'e Paris-Sud, CNRS-UMR 8502, Orsay 91405, France}

\author{Jean-Denis~Blazit}
\affiliation{Laboratoire de Physique des Solides, Universit\'e Paris-Sud, CNRS-UMR 8502, Orsay 91405, France}
\author{Xiaoyan~Li}
\affiliation{Laboratoire de Physique des Solides, Universit\'e Paris-Sud, CNRS-UMR 8502, Orsay 91405, France}
\author{Alexandre~Gloter}
\affiliation{Laboratoire de Physique des Solides, Universit\'e Paris-Sud, CNRS-UMR 8502, Orsay 91405, France}
\author{Alberto~Zobelli}
\affiliation{Laboratoire de Physique des Solides, Universit\'e Paris-Sud, CNRS-UMR 8502, Orsay 91405, France}

\author{Luis~Liz-Marz\'an}
\affiliation{CIC biomaGUNE and Ciber-BBN, Paseo de Miram\'on 182, 20014 Donostia-San Sebasti\'an, Spain}
\affiliation{Ikerbasque, Basque Foundation for Science, 48013 Bilbao, Spain}

\author{F.~Javier~Garc\'{\i}a~de~Abajo}
\affiliation{ICFO-Institut de Ciencies Fotoniques, The Barcelona Institute of Science and Technology, 08860 Castelldefels (Barcelona), Spain}
\affiliation{ICREA-Institució Catalana de Recerca i Estudis Avançats, Passeig Llu\'is Companys 23, 08010 Barcelona, Spain}

\author{Odile~St\'ephan}
\affiliation{Laboratoire de Physique des Solides, Universit\'e Paris-Sud, CNRS-UMR 8502, Orsay 91405, France}
\author{Mathieu~Kociak}
\email{mathieu.kociak@u-psud.fr}
\affiliation{Laboratoire de Physique des Solides, Universit\'e Paris-Sud, CNRS-UMR 8502, Orsay 91405, France}

\date{\today}

\begin{abstract}
Vibrational optical spectroscopies can be enhanced by surface plasmons to reach molecular-sized limits of detection and characterization. The level of enhancement strongly depends on microscopic details of the sample that are generally missed by macroscopic techniques. Here we investigate phonons in \textit{h}-BN by coupling them to silver-nanowire plasmons, whose energy is tuned by modifying the nanowire length. Specifically, we use electron beam milling to accurately and iteratively change the nanowire length, followed by electron energy-loss spectroscopy to reveal the plasmon-enhanced vibrational features of \textit{h}-BN. This allows us to investigate otherwise hidden bulk phonons and observe strong plasmon-phonon coupling. The new milling-and-spectroscopy technique holds great potential for resolving vibrational features in material nanostructures.

\end{abstract}

%\keywords
\maketitle
Vibrational spectroscopy is key in a wide range of research areas and technological applications, from molecular fingerprinting to fundamental solid-state physics \cite{Z17,V17}. The discovery that plasmonic structures can increase the measured vibrational signal has driven the development of ultra-sensitive analytical techniques capable of reaching single-molecule detection, such as in surface-enhanced Raman spectroscopy \cite{FHM1974,NE97} (SERS) and surface-enhanced infrared absorption \cite{OI91} (SEIRA). Typically, plasmonic structures are designed in advance and molecules are randomly dispersed over them, leading to strongly enhanced signals associated with those sitting on the so-called hotspots\cite{NE97}. Alternatively, a metallic tip can be scanned over a sample to induce SERS locally, leading to chemical mapping at sub-molecular scales, a technique known as tip-enhanced Raman scattering \cite{SPJ14} (TERS). 

Vibrational electron energy-loss spectroscopy (EELS) has been performed for decades mainly as a surface technique \cite{IM1982} using wide beams with poor spatial resolution. The recent development of a new family of electron monochromators \cite{KUB09} has allowed vibration mode measurements to be performed based on EELS \cite{KLD14,LTHB17,SSB18}  down to atomic spatial resolution in bulk materials\cite{HKR19}.
Unfortunately, the signal-to-noise ratio of vibrational EELS is low for materials that are sensitive to the electron beam, thus imposing a lower bound on the volume necessary for analysis. This jeopardizes the high resolution mapping of fragile materials, such as organic molecules \cite{RAM16}. Recently, theoretical studies \cite{NPC08,KNA18,KD19} have proposed the use of infrared plasmonic fields to develop a new form of enhanced vibrational EELS which would overcome these limitations by making the molecules interact with the beam at a distance mediated by plasmons extended in a nanoparticle.

\begin{figure*}
\includegraphics[width=12cm]{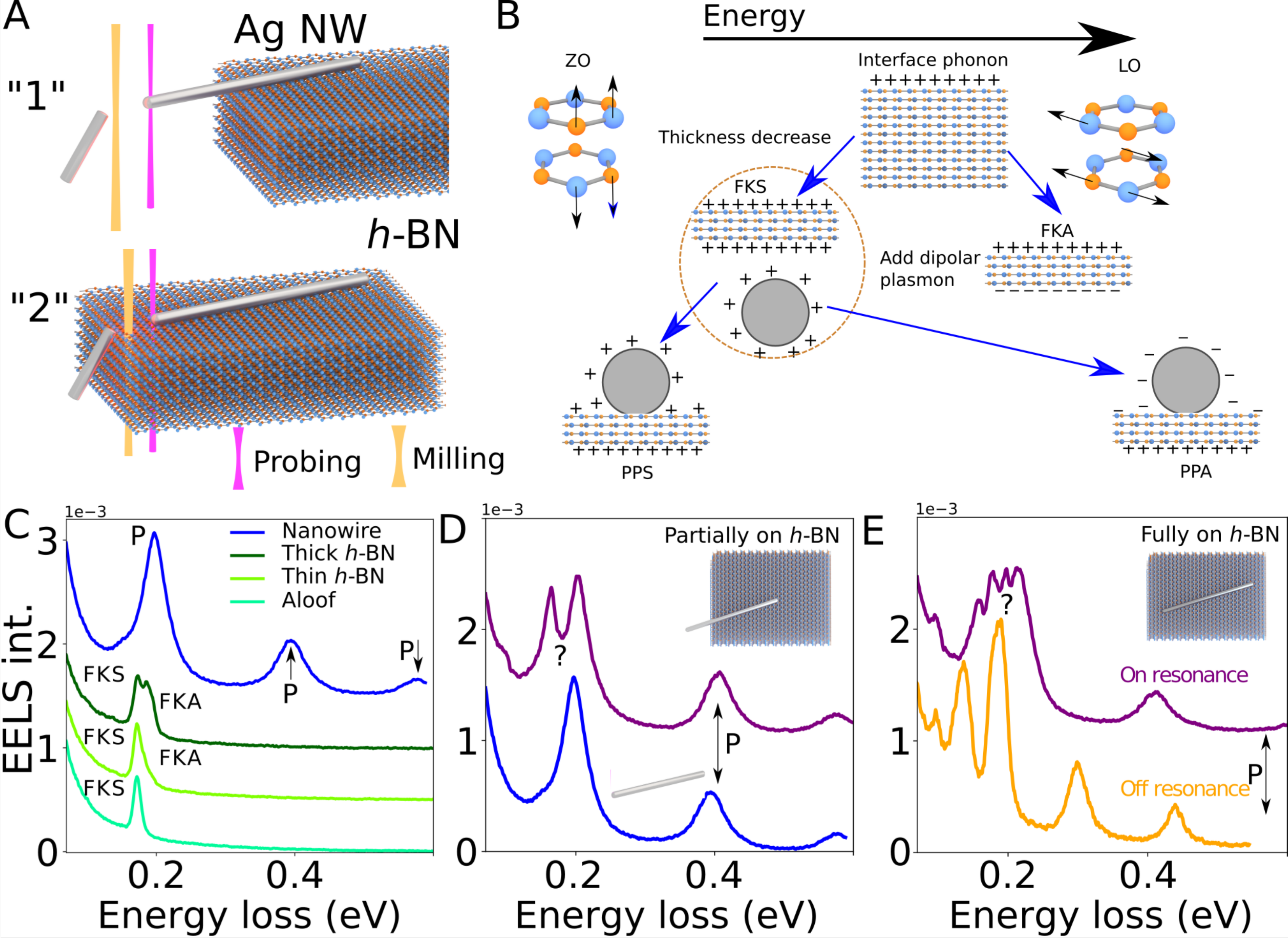}
\caption{\textbf{(A)} Samples consisting of silver nanowires deposited on \textit{h}-BN are investigated with two different configurations:  (1) nanowires partially supported by \textit{h}-BN with one of its ends in vacuum; and (2) nanowires fully supported on \textit{h}-BN. High-current ($\sim10\,\mu$A at 200\,keV, orange) and low-current ($\sim10\,$pA at 60\,keV, purple) electron beams are used for milling and probing, respectively. \textbf{(B)} Sketches of the different phonons, plasmons, and hybrid excitations discussed in the text. \textbf{(C)} Spectra of a 3.3\,$\mu$m long metallic nanowire in vacuum (blue) and \textit{h}-BN (green). The in-vacuum nanowire exhibits equally spaced spectral modes labeled P. \textit{h}-BN spectra are shown for the electron beam placed outside the material (aloof configuration) and on either thin or thick \textit{h}-BN, where the symmetric (FKS) or antisymmetric (FKA) Fuchs-Kliever modes are observed.  \textbf{(D)} Spectrum of a metallic nanowire in sample configuration (1) measured on the tip in vacuum (upper curve). The dipolar mode at $\sim200\,$meV (black arrow) is split into two peaks compared with the nanowire in vacuum (lower curve).  \textbf{(E)} Spectra for the same metallic nanowire in configuration (2)  under off- (orange) and on-resonance (purple) conditions.  \label{Figure1}}
\end{figure*}

Here, we demonstrate plasmon-enhanced vibrational electron spectroscopy (PEVES) through the tailored coupling of plasmon resonances in metallic nanowires to phonon modes in \textit{h}-BN thin flakes. Coupling is achieved by continuously shifting the energies of the plasmon modes of micrometer-long metallic nanowires (Fig. S1) using electron-beam controlled milling to bring them into resonance with specific \textit{h}-BN vibrational modes (Fig. S2).  We reveal three new effects when a plasmon-phonon resonance is encountered: 1) strong coupling between surface phonons and plasmons; 2) enhancement of the bulk vibrational EELS signal; and 3) emergence of previously geometry-forbidden dark phonon modes. 

EELS measures the distribution of energy losses experienced by free electrons when interacting with a target \cite{E96}. The energy resolution is determined by the energy spread of the electron source, typically $\sim$250\,meV for a cold field emitter source. This resolution can be improved down to a few meV using an electron monochromator (Fig.\ S3). To achieve subnanometer spatial resolution, the electron monochromator can be coupled to an electron microscope (Fig.\ S3). Finally, an electron spectrometer is used to acquire the spectrum of the electron beam after interaction with the sample. Here, we use such a setup implemented on a NION Hermes scanning transmission microscope (STEM), see Methods.

The plasmonic silver metallic nanowires were synthesized by chemical seeded growth as detailed elsewhere \cite{paper258} (Fig.\ \ref{Figure1}A) and subsequently deposited, either entirely (configuration (1)) or partially (configuration (2)), on an \textit{h}-BN substrate \cite{T2007}. 
In theory, extended \textit{h}-BN possesses a variety of infrared-active either transverse optical (TO) or longitudinal optical (LO) phonons. Due to the \textit{h}-BN anisotropy, TO and LO phonons may possess different energies depending on whether they are polarized out- or in-plane \cite{GPR1966}. Bulk EELS is therefore very dependent on the orientation of the electron to the \textit{h}-BN crystallographic orientation. Indeed, when the electron propagates along the [0001] \textit{h}-BN direction, mainly the in-plane LO bulk mode is excited, while the out-of-plane modes can only be excited in a tilted configuration. Thin \textit{h}-BN substrates sustain additionally two surface phonon modes: the charge-symmetric (FKS) and charge-antisymmetric (FKA) Fuchs-Kliewer  modes, whose energies disperse as a function of thickness. A qualitative sketch of the energy hierarchy of the different modes discussed in the text is presented in Fig. \ref{Figure1}B. For very thin substrates, the former (FKS) arises at the in-plane TO phonon energy (around 169 meV) and the later (FKA) at the out-of-plane LO phonon energy (around 198 meV). In EELS experiments,  the FKS (resp. FKA) energy varies between the TO (resp. LO) energy and that of the \textit{h}-BN interface phonon (around 195 meV). Also, the FKA intensity is usually much smaller than that of the FKS \cite{batson2017,LK17}, see (Fig.\ \ref{Figure1}C, green curves). The FK modes usually overshadow the bulk modes, at least for the thickness range probed in this study (50-100 nm). Nanowire plasmons of well-defined energy are produced due to confinement, leading to modes that can be classified as dipolar, quadrupolar, etc., depending on the number of nodes in the induced-charge along the nanowire length. In the nanowire cross-section, the charge distribution is homogeneous. EELS is extremely sensitive to surface plasmon modes and reveals their nearly uniform energy spacing (Fig.\ \ref{Figure1}C). Their energies are roughly proportional to the inverse of the nanowire length, which thus becomes a perfect parameter to tune the nanowire plasmon energies. Finally, we note that both surface plasmons and surface phonons (FK modes) can also be excited in a geometry where the beam does not directly hit the sample ("aloof geometry") \cite{paper085,KLD14,LTHB17}.

\begin{figure*}
\includegraphics[width=12cm]{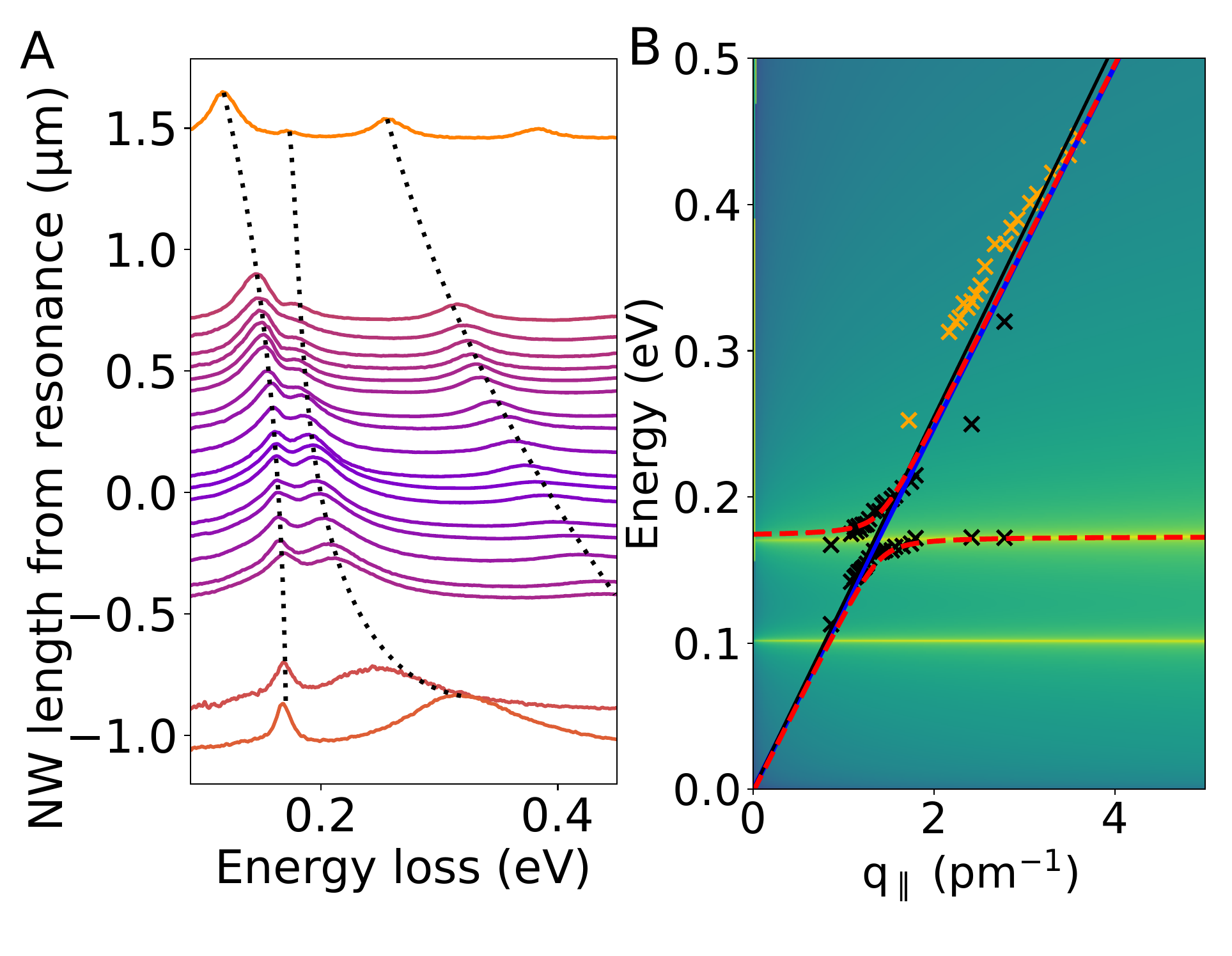}
\caption{\textbf{Plasmon-phonon strong coupling} \textbf{(A)} EELS spectra measured at the tip in vacuum of a metallic nanowire (under sample configuration (1) in Fig.\ \ref{Figure1}A) as a function of nanowire length. By changing through milling the length of the nanowire, its dipolar plasmon mode is brought in and out strong coupling with the \textit{h}-BN phonon modes at around 180 meV. The strong coupling is confirmed by the anti-crossing of the two modes, indicated by the two dashed curves (added as guides to the eye) on the left of (A). \textbf{(B)} Theoretical calculations for the energies of the coupled modes (dashed red curve) and the energy position measured in experiments (black and orange crosses for the dipolar and quadrupolar modes). The light line is marked in blue. The quadrupolar plasmon mode is marked by orange crosses. \label{Figure2}}
\end{figure*}

Plasmon-phonon interaction was probed initially using a sample configuration ("1" in Fig. \ref{Figure1}A) where only surface phonons interact with surface plasmons, to avoid any potential overlap between surface and bulk effects. In this geometry, long silver nanowires with only one tip placed on different volumes of \textit{h}-BN were modified in length by electron milling in order to bring their dipolar plasmon into and then out of resonance with respect to the \textit{h}-BN FK modes.

When their energies differ by more than $\sim100\,$meV, phonons and surface plasmons do not interact (Figs.\ \ref{Figure1}D, blue curve, and \ref{Figure2}A, orange curves). As the nanowire length is shortened, the dipolar plasmon energy approaches that of the FKS mode. Close to resonance, a second peak is observed at the nanowire tip in vacuum, away from \textit{h}-BN, at an energy close to 180\,meV (Figs.\ \ref{Figure1}D and \ref{Figure2}A, purple curves). The two peaks do not cross each other as the nanowire length is varied, exhibiting a minimum splitting of 37\,meV (average splitting in the $1.7$-$2.9\,\mu$m nanowire length range), comparable to the full width at half maximum of the two peaks (between 30 and 40\,meV). The energy of the first peak continues to increase for nanowire lengths shorter than $2.9\, \mu$m. However, it does not disperse proportionally to the inverse of the nanowire length $1/L$, as expected for plasmons on long metallic nanowires. This is in contrast with the higher order plasmon modes that indeed disperse as expected as $1/L$ (Fig.\ S4). This behavior is nicely reproduced by numerical calculations (Fig.\ \ref{Figure2}B). Peak-splitting was observed to vary between nanowires (in the 0-52 meV range in Figs. S5-8).

These observations indicate strong coupling between the dipolar surface plasmon and the FKS mode. For small detuning, the two peaks cannot be assigned as either a plasmon or a phonon, but rather as hybrid plasmon-phonon (PP) modes, either charge-symmetric (low-energy, PPS) or charge-antisymmetric (high-energy, PPA), see Fig.\ \ref{Figure1}B. After the anticrossing, the nature of the modes is inverted, as expected for strong coupling: the lower (higher) energy peak changes continuously from the plasmon (phonon) to the phonon (plasmon) mode.  Strong coupling evidences a coherent field across the whole nanowire, as its signature is observed at a region of the nanowire where no \textit{h}-BN is present and the aloof signal from \textit{h}-BN without nanowire is not experimentally detectable. In fact, the coupled peaks are only observed where the dipolar mode is measurable, close to the nanowire tips (Fig.\ S9). This can be more clearly observed in a line profile of the hyperspectral image going from the nanowire tip in vacuum to the region supported on thin \textit{h}-BN (Fig.\ S9). As the electron beam is moved away from the nanowire tip, the two coupled modes lose intensity. When it approaches \textit{h}-BN, the lower energy FKS mode is observed at an intermediate energy between the two coupled modes. Finally, with the electron beam on \textit{h}-BN, both FK modes are revealed. 

The energy splitting due to coupling depends on the amount of \textit{h}-BN contained inside the modal volume of the dipole surface plasmon \cite{KNA18}. For small quantities, no splitting is observed (Fig.\ S6). For intermediate amounts, a 17\,meV splitting is observed in some measurements (Fig.\ S7). The largest observed splitting in this study is 52\,meV Fig. \ S8. These results support the hypothesis that the volume of \textit{h}-BN within the dipole surface plasmon modal volume determines the splitting of the two strongly-coupled peaks, starting from a large volume of \textit{h}-BN, where the splitting continuously decreases from 52\,meV down to 12\,meV (Fig.\ S8). We note that a removal of \textit{h}-BN  performed only within 500 nm of the nanowire tip did modify the splitting, while removal farther away did not affect the splitting, again indicating that the coupling is mediated by the overlap of the plasmon field lines and \textit{h}-BN.

\begin{figure*}
\includegraphics[width=16cm]{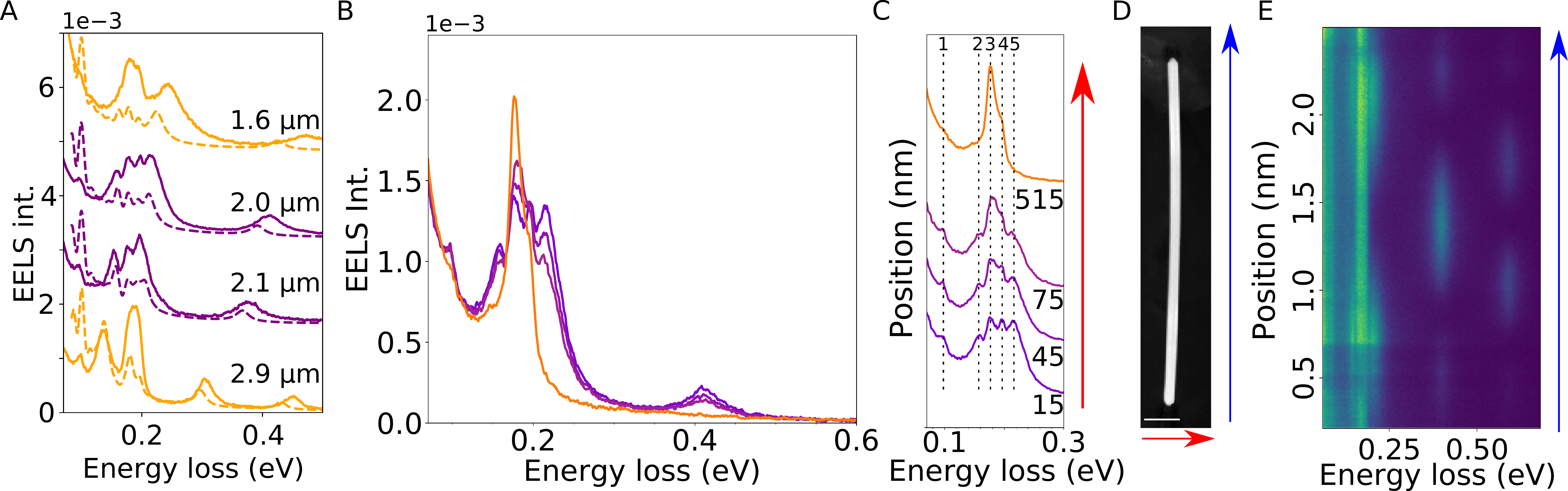}
\caption{\textbf{Plasmon enhanced phonon losses} \textbf{(A)} Experimental (full curves) and calculated (dashed curves) EELS spectra at one of the tips of a metallic nanowire fully supported on \textit{h}-BN (configuration 2 in Fig.\ref{Figure1}A) for different lengths of the nanowire: 1.6 , 2.0, 2.1, and 2.9\,$\mu$m. When the dipolar plasmon mode is close to or on resonance with \textit{h}-BN phonon modes, the latter are enhanced, leading to the appearance of multiple peaks.\textbf{(B-C)} Spatial evolution of the EELS spectra along a direction perpendicular to the nanowire close to the bottom tip (red arrow marked  in (D)). The ZO, PPS, FKS, LO, and PPA modes are marked by 1, 2, 3, 4, and 5, respectively in (C) \textbf{(D)} STEM image of the nanowire on \textit{h}-BN. \textbf{(E)} Spatial evolution of the EELS spectra intensity as a function of energy loss and position along the nanowire length (blue arrow in (D)).   \label{Figure3}}
\end{figure*}

The above-described behavior is observed at the nanowire tip lying on \textit{h}-BN, but with the complicating factor that \textit{h}-BN losses are also present. To understand the signal when \textit{h}-BN is present, a second sample geometry ("2" in Fig. \ref{Figure1}A), in which the nanowire is lying completely on an homogeneous piece of \textit{h}-BN, was used.
The presence of plasmons with energies away from the phonon energies (Fig.\ \ref{Figure1}E, orange curve) does not affect the phonon spectra. Close to a plasmon-phonon resonance, the spectra change substantially (Fig.\ \ref{Figure1}E, purple curve and S10), with the appearance of multiple peaks at 100 meV, 161 meV, 177 meV, 197 meV, and 217 meV  (Fig..\ \ref{Figure3}A, B and marked 1 through 5, respectively in C). This multiple peak structure  is reproduced by theoretical calculations (dashed curves in Fig.\ \ref{Figure3}A). The 161 meV and 217 meV modes do disperse with nanowire length, and can be safely attributed to the PPS and PPA modes (their dispersion is confirmed by numerical calculations shown in Fig. S11). Three peaks are left to be identified (100, 177, and 197 meV). For the 2.1\,$\mu$m long nanowire, peak energies do not change as a function of the distance away from the nanowire tip at which they are probed (Fig. \ref{Figure3}B-C and S12), but their intensities do. 

 The 177 meV mode intensity increases away from the nanowire surface, while the four others vanish (Fig. \ref{Figure3}B-C). The 177 meV mode can be interpreted as the FKS, excited in an aloof-like configuration when the beam is close to the nanowire tip. The probability of exciting it obviously decreases as the PPA and PPS, which are hybrid plasmon-FKS modes, are more intensely excited close to the nanowire tip.

Away from the influence area of the dipolar mode, the vanishing of the PPA and PPS signals is expected, as they are formed where the plasmon field lines are intense enough. The 100 and 197 meV peaks show similar spatial variation, indicating that they have a common origin. The first of these energies matches that of the ZO normal mode \cite{SBA07} (a longitudinal optical phonon polarized along the anisotropy axis), and the second one corresponds well with that of the LO mode polarized perpendicular to the anisotropy axis (see a schematics on Fig. \ref{Figure1}B). As these modes should not be observed in the sample geometry here considered, their observation can only be explained by their coupling to the plasmon mode, which enhances them. The enhancement is maximal within 30 nm of the nanowire surface, which strengthens the link to the surface plasmon.

In summary, two main effects are observed: the possibility to detect a bulk phonon mode (the ZO mode), whose excitation, in the absence of a nanowire, is geometry-forbidden, and an extreme enhancement of the LO and ZO modes. The magnitude of the enhancement cannot be quantified, as in the absence of nanowire both modes are not experimentally detectable.

 Interestingly, the enhancement is not homogeneous across the whole nanowire: both ZO and LO modes are enhanced at the tips, while only the ZO mode is enhanced at the center (Fig.\ \ref{Figure3}D,E). We attribute this effect to the dependence of the enhancement on the direction of the plasmon electric field, which changes along the nanowire length: close to the tips the field has substantial projections on all directions, while at the nanowire center it is mostly oriented parallel to the metal surface. Some phonon modes are only observed when the electric field has a component out of the \textit{h}-BN plane. The PPS and PPA peaks are visible close to the nanowire tips (left and right of Fig.\ \ref{Figure3}D), while others appear along the entire nanowire (Fig.\ \ref{Figure3}D, ZO peak). In this respect, the symmetry of the plasmon electric field mode provides additional information about the phonons with which it interacts.

Controlled coupling between plasmon and phonon modes with precision below 10\,meV opens the way to a new form of plasmon-enhanced vibrational spectroscopy (PEVES). Plasmon modes are highly localized in space, leading to stronger enhancement effects in volumes of the order of $10^3$\,nm$^3 $  \cite{NPC08}; only a technique with sub-10\,nm spatial resolution can exploit this high localization. The plasmonic enhancement has the additional benefit that smaller sample volumes can be probed than with regular EELS. More advanced sample designs should allow fingerprinting of molecular vibration modes. For example, one could produce dedicated slots in metallic structures, each of them containing different analytes; or a movable metallic tip could be added to the microscope holder (as demonstrated in previous work\cite{OKT98}) to selectively enhance the vibrational modes along a molecule. Sequencing of DNA strands also appears as a possible application of PEVES by scanning both the tip and the electron beam.

\begin{acknowledgments}

This project has received funding from the european union’s horizon 2020 research and innovation programme under grant agreement no 823717.This work has received support from the National Agency for Research under the program of future investment TEMPOS-CHROMATEM with the Reference No. ANR-10-EQPX-50. We acknowledge support from Spanish MINECO (MAT2017-88492-R and SEV2015-0522), ERC (Advanced Grants 789104-eNANO and 4DBIOSERS, AdG 787510), the Catalan CERCA Program, and Fundaci\'o Privada Cellex.
The authors acknowledge Christoph Hanske for help in the nanowire sample preparation.
\end{acknowledgments}

\bibliography{hBNPlasmonCoupling.bib}

\pagebreak
\widetext
\begin{center}
\textbf{\large Supplemental Materials: Tailored nanoscale plasmon enhanced vibrational electron spectroscopy}
\end{center}
%%%%%%%%%% Merge with supplemental materials %%%%%%%%%%
%%%%%%%%%% Prefix a "S" to all equations, figures, tables and reset the counter %%%%%%%%%%
\setcounter{equation}{0}
\setcounter{figure}{0}
\setcounter{table}{0}
\setcounter{page}{1}
\makeatletter

%%%%%%%%%% Prefix a "S" to all equations, figures, tables and reset the counter %%%%%%%%%%

\textbf{This PDF file includes:}

Materials and Methods

Figs. S1 to S12

\section{Materials and Methods}
\subsection{Electron spectroscopy and microscopy}

Experiments were performed on a modified Nion Hermes200 (known as ChromaTEM) equipped with an electron monochromator, an optimized electron spectrometer (Nion Iris) and a cold field emission electron gun. The NION Iris spectrometer was fitted with a Princeton Instrument Kuros camera. The electron beam used had a kinetic energy of 60 keV, an energy resolution (defined as the full width at half maximum (FWHM) of the zero-loss peak) between 6 and 10 meV (this value was optimized and controlled after each acquisition), a current intensity between 5 and 20 pA, a convergence half-angle of 10 mrad, and a FWHM in real space of about 1 nm. The spectrometer was set to have a diffraction pattern at its entrance (image coupling mode) and a 1 mm entrance aperture was used to collect electrons up to 15 mrad (half-angle). All data were acquired as hyperspectral images, with two spatial dimensions (electron beam scan in real space) and one energy loss dimension. 
All peak positions have been reported with an error of $\pm$ 5 meV (which include errors due to the background subtraction).

\subsection{Sample preparation}

The synthesis follows the protocol reported by Mayer et al. in 2015 \cite{paper258}. In order to proceed with the growth of AgAuAg bimetallic nanowires, it is necessary to synthetize Penta-twinned gold nanorods to be used as core for the epitaxial deposition of silver. The synthesis of this material can be found in the literature \cite{S2016}.

\textbf{Materials}. Benzyldimethylhexadecylammonium chloride (BDAC), silver nitrate (AgNO3, $\geq 99.9 \%$), L-ascorbic acid (AA, $\geq 99 \%$), were purchased from Aldrich. All chemicals were used as received. Milli-Q water (resistivity 18.2 M$\Omega$cm at 25 $^\circ$C) was used in all experiments. All glassware was washed with aqua regia, rinsed with water, and dried before use.

\textbf{AgAuAg bimetallica nanowires synthesis}. 20 mL of the purified Penta-Twinned Au Nanorod solution was heated to 60 $^\circ$C using an oil bath. A AgNO3 solution (0.004 M in water) and an AA solution (0.016 M in 20 mM BDAC) were prepared and loaded into separated syringes. Using a syringe pump, both AgNO3 and AA were added to the Penta-Twinned Au Nanorod solution at a rate of 300 $\mu$L/h (0.24 mol of Ag(I) per mol of Au(0) per hour) under slow stirring keeping the temperature constant at 60 $^\circ$C. The growth of the nanowires was monitored through UV-vis-NIR spectroscopy on a small sample (1 mL) periodically withdrawn from the growing solution. To obtain AgAuAg nanowires above 3 $\mu$m in length (average 3.4 $\pm$ 0.6 $\mu$m for the specific sample used in this manuscript), the growth was continued for 72 hours. 

\subsection{AgAuAg nanowire and \textit{h}-BN milling}

Metallic nanowire and \textit{h}-BN milling was performed on a JEOL 2010 electron microscope equipped with a thermionic electron gun (LaB$_6$ crystal). This kind of microscope was chosen to ensure the maximum total current available in the focused electron beam (in contrast to the maximum current density, which would be achieved with cold field emission gun). Most efficient milling was achieved with 200 keV kinetic energy electrons and a beam extraction current between 10 and 15 $\mu$A.

Milling of the two materials was achieved through different mechanisms. Ag atoms are heavy and energy conservation prevents efficient atoms removal by momentum transfer. For this reason, large currents were used to heat the material and diffuse atoms away from the target area. Because of this, milling was quite dependent on the underlying substrate. The highest precision was achieved with nanowires parts on vacuum or on \textit{h}-BN. On the carbon support film milling was less favorable, probably due to good heat conductivity of the film.
B and N atoms are light and are easily removed by direct momentum transfer from the electron beam.     

\subsection{Data analysis}
Data acquisition and initial check was performed using Nion Swift and in-house Kuros driver. Data analysis was performed with Python libraries: Hyperspy, Numpy, Matplotlib and Scipy. Spectra were fitted with multiple Gaussians unless otherwise noted. 

\subsection{Sample preparation}
Samples were prepared by drop casting sequentially solutions containing the metallic nanowires and \textit{h}-BN flakes (chemically exfoliated in IPA). Nanowires where chosen by inspection in an electron microscope considering their length and the quantity of \textit{h}-BN underlaying them.

\subsection{Purcell factor calculations}
In order to understand the experimental measurements more quantitatively, we present numerical simulations of the optical response of our system (Ag nanowires on top of \textit{h}-BN thin films) carried out by using a finite-difference in the time domain (FDTD) method. In these simulations, plasmon modes are excited by an electric dipole source placed near the end of the nanowire, at a position away from high-symmetry points in order to guarantee a nonzero electric field of the plasmon standing wave, and therefore, a high excitation efficiently.
We exploit the proportionality between the EELS probability and the local density of electromagnetic states (LDOS) [1]
\begin{equation}
    \Gamma_{EELS} = \frac{-2\pi e^2}{\hbar\omega}\rho_z(R_0, q_v, \omega)
\end{equation}

where $e$ is the elementary charge, $\omega$ is the frequency of the electromagnetic field, $q_v=\omega/v$ , and $v$ is the electron velocity. Here, $\rho_z(R_0, q_v, \omega)$ is the projected LDOS (along the direction of the electron trajectory, which is taken along the $z$ axis) at the in-plane position $ R_0(x_0,y_0)$  of the electron trajectory and evaluated in wave vector space along the $z$ direction. Instead of this quantity, we compute the LDOS in real space along $z$, which is in turn proportional to the so-called the Purcell factor (i.e., the enhancement of the decay rate  of a dipole emitter placed at the position of interest, compared with the free space decay rate $\gamma_0$). More precisely [2],

\begin{equation}
    P = \frac{\gamma}{\gamma_0} = \frac{2\pi c^2}{\omega^2}\rho_z (R_0, z, \omega)
\end{equation}

Although the LDOS forms in two expressions above are not equivalent, we argue that both plasmons and polaritons are extremely localized in space, so the main contribution to EELS will be roughly described by the second form, which is more direct to compute in our method. 

\subsection{Strong coupling}

Experimental data shows the anti-crossing behavior apparent to the strong coupling regime between plasmon and phonon modes. This can be modeled using a classical model of strongly coupled harmonic oscillators whose dispersion is give by a simple expression

\begin{equation}
    \frac{1}{2}(\omega_{pl}+\omega_{ph})\pm(\frac{1}{4}(\omega_{pl}-\omega_{ph})^2+g^2)^{\frac{1}{2}}
\end{equation}

where $\omega_{pl}$ is the plasmon resonant frequency, $\omega_{ph}$ is the phonon resonant frequency and $g$ is the coupling strength between plasmons and the phonons. The later served as a fitting parameter for us to fit the experimental data using the analytically calculated plasmon $\omega_{pl}(q)$  dispersion of the infinite nanowire and the phonon  dispersion $\omega_{ph}(q)$  curves for the h-BN thin film. 
The infinite nanowire plasmon dispersion can be shown to be given with the expression\cite{AE1974}
\begin{equation}
    \frac{m q^2}{a^2}\left( \frac{1}{k^2_{2\perp}} - \frac{1}{k^2_{1\perp}}\right)^2 = \left[ \frac{1}{k_{2\perp}} \frac{J^{'}_m (k_{2\perp}a)}{J_m(k_{2\perp}a)} - \frac{1}{k_{1\perp}} \frac{H^{(1)'}_m (k_{1\perp}a)}{H^{(1)}_m(k_{1\perp}a)} \right] x \left[ \frac{k^2_2}{k_{2\perp}} \frac{J^{'}_m (k_{2\perp}a)}{J_m(k_{2\perp}a)} - \frac{k^2_1}{k_{1\perp}} \frac{H^{(1)'}_m (k_{1\perp}a)}{H^{(1)}_m(k_{1\perp}a)} \right]
\end{equation}

where $a$ is the wire radius, $m$ is the azimuthal order of the plasmon ($m=0$ in our case), $k_j=\epsilon_j k_0$ ,$k_0=\omega/c$ , and $k_{j\perp} = \sqrt{k^2_j-q^2}$ is the transverse wave vector in medium $j$. The phonon-polariton modes of the \textit{h}-BN film are calculated from Fresnel reflection coefficient. In particular, for Fig. S13A we use the imaginary part of the reflection coefficient $\Im(r_p)$for the $p$ polarized fields , which readily gives the hyperbolic phonon-polaritonic bands (KF modes). In Fig. S13B we compare the coupling strength with the plasmon and phonon resonance widths that are again extracted from simulations. It is clear that the coupling strength is larger than the plasmon and phonon widths for small values of $q$, therefore indicating that the system undergoes strong coupling

\pagebreak
\section{Supplementary figures}

\begin{figure*}[h]
\includegraphics[width=16cm]{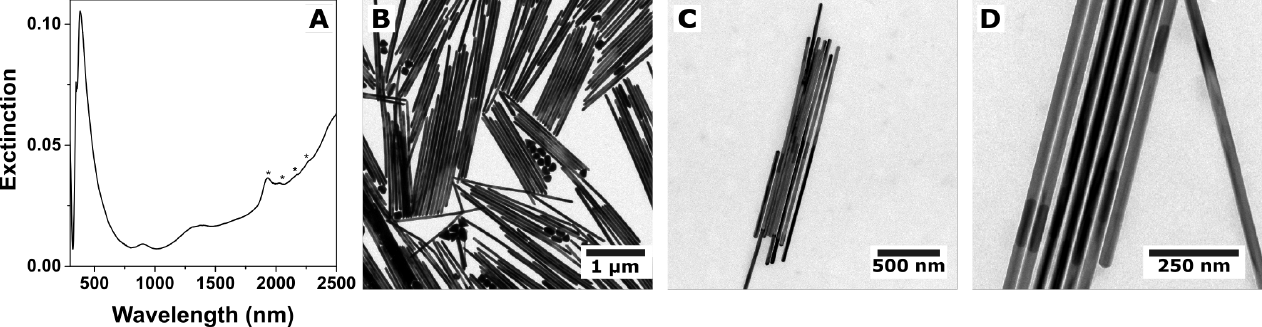}
\caption{\textbf{Characterization of AgAuAg bimetallic nanowires.} \textbf{(A)} UV-vis-NIR spectra of the nanowire colloidal suspension used in this study, taken in deuterated water. Asterisks indicate residual water peaks. \textbf{(B-D)} TEM images taken at different filed of views. In panel D the gold cores are clearly visible.}
\end{figure*}

\begin{figure*}
\includegraphics[width=16cm]{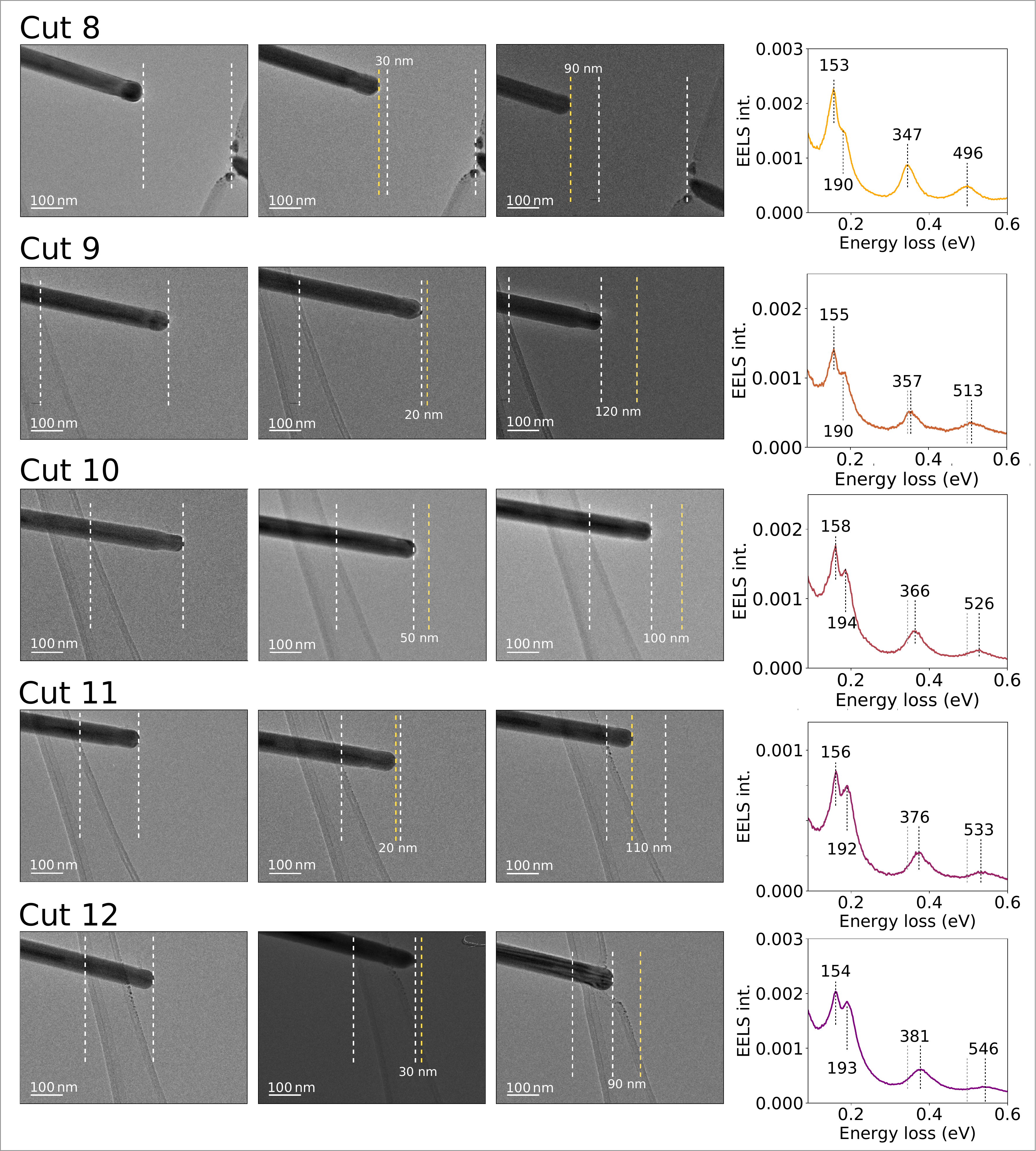}
\caption{\textbf{Plasmon tuning by nanowire shape changes.} The three panels on the left of each line show TEM images at initial (lef), intermediary (middle) and final (right) steps of cuts 8, 9, 10, 11 and 12 of the nanowire presented in Fig. 2 of the main text. White lines are guides to the eyes. In the right panel, corresponding spectra measured at the tip are shown. Their energy positions are displayed in milli-electronvolt and were extracted from multi-gaussian curve adjustment to the data. Dotted lines are guide to the eyes and grey lines mark their original position in cut 8. The quadrupolar and sextupolar modes (initially at 347 and 496 meV) disperse to higher energy as expected for plasmon modes. The two strong-coupled modes (initially at 153 and 190 meV) do not disperse within our precision in the length range.}
\end{figure*}

\begin{figure*}
\includegraphics[width=16cm]{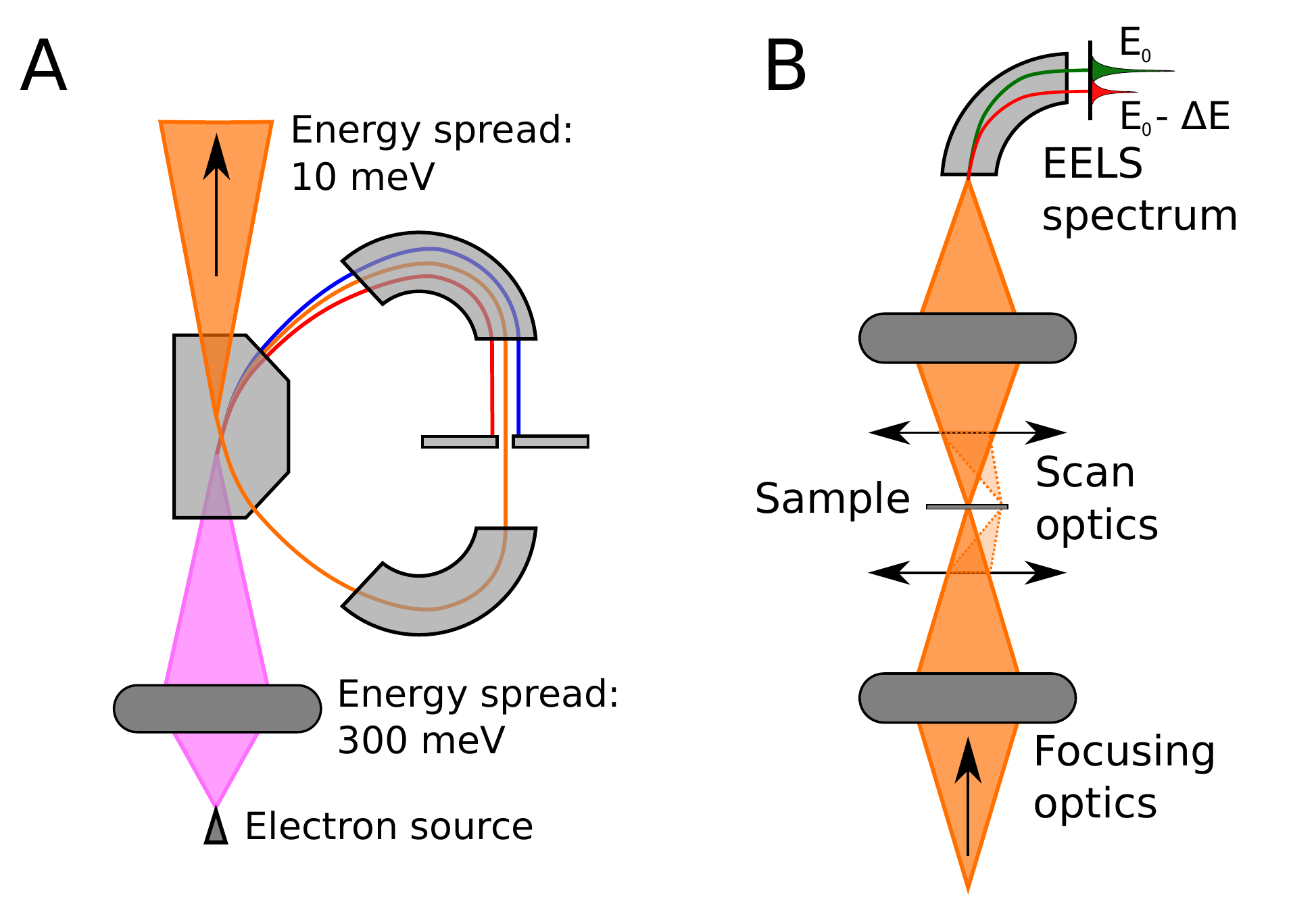}
\caption{\textbf{General description of the experimental configuration.} \textbf{(A)} An electron  monochromator is used to reduce the energy width of a free electron beam down to 5-10 meV. \textbf{(B)} An electron microscope focuses the monochromated electron beam down to subnanometer sizes. The focused beam is scanned in real space to construct energy-resolved maps of the samples. }
\end{figure*}

\begin{figure*}
\includegraphics[width=16cm]{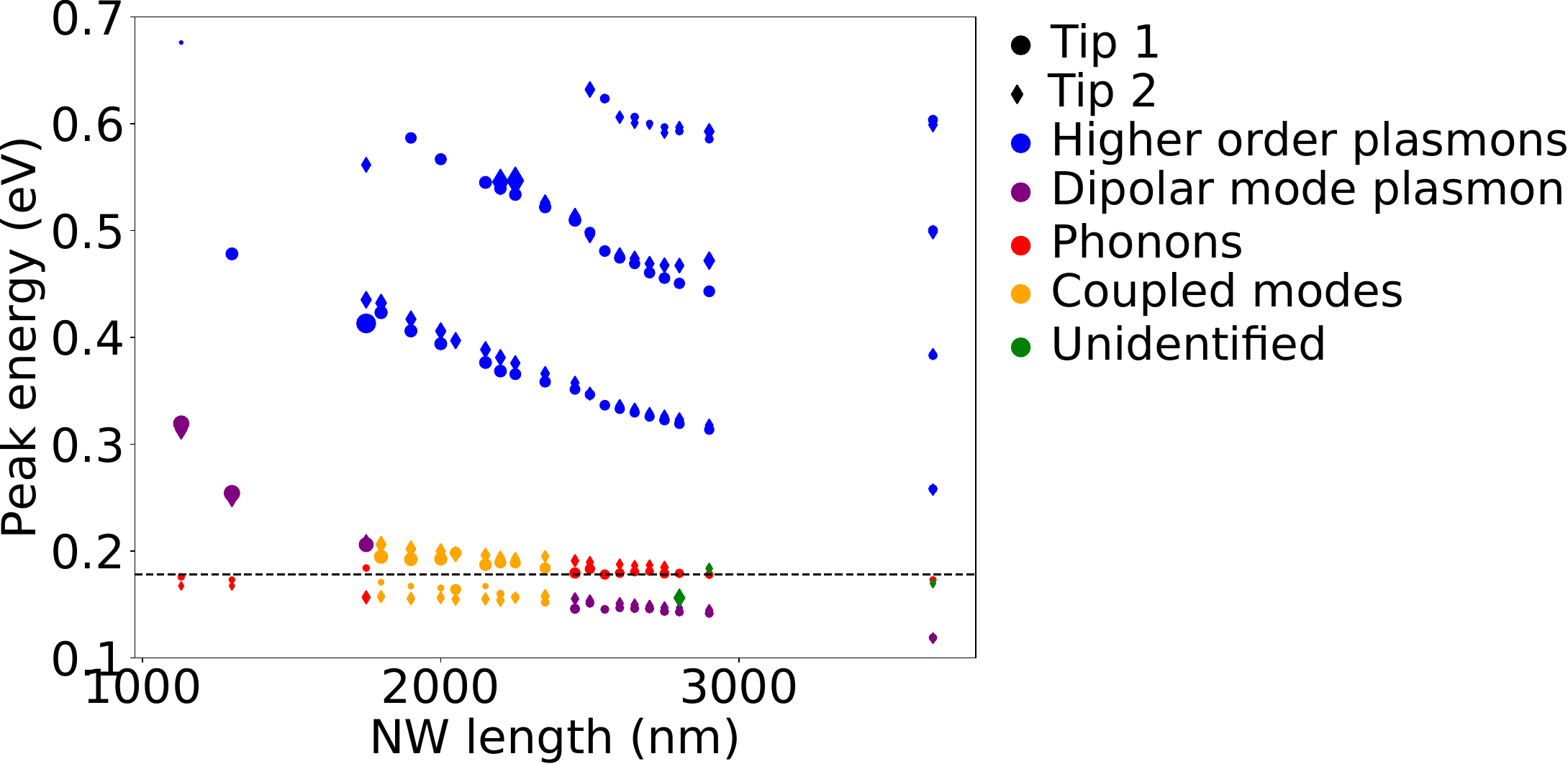}
\caption{\textbf{Dispersion of the higher order plasmon modes versus energy.} The distinction between dipolar plasmon mode/phonons and coupled modes is somewhat arbitrary and given as a guide to the eyes.}
\end{figure*}

\begin{figure*}
\includegraphics[width=9cm]{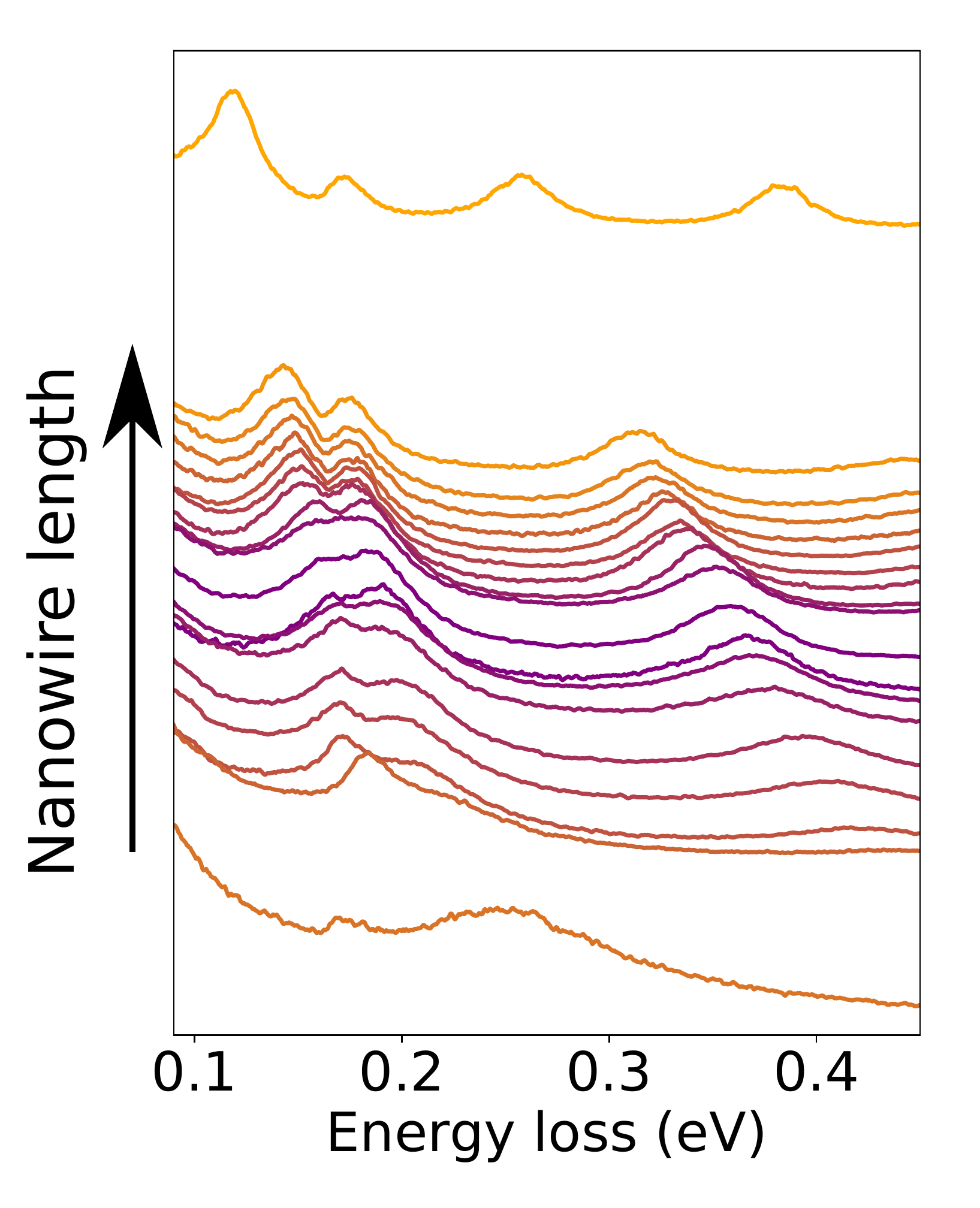}
\caption{\textbf{Spectra measured on tip 1 (on \textit{h}-BN) of the nanowire presented in Fig. 2 of the main text.}}
\end{figure*}

\begin{figure*}
\includegraphics[width=12cm]{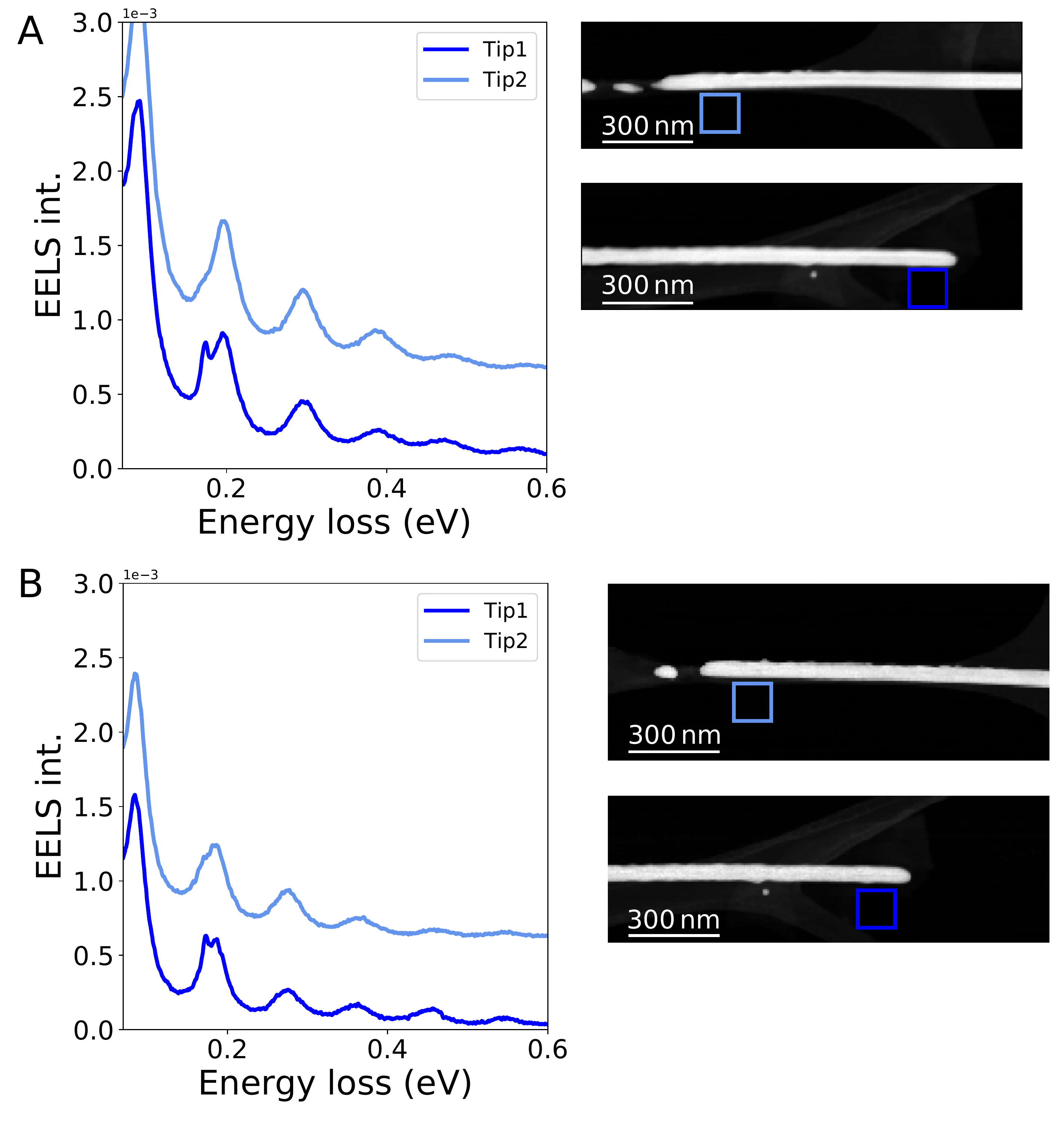}
\caption{\textbf{Nanowire with small amount of \textit{}{h}-BN and no coupling.} (A) and (B) show EELS spectra measured at tips 1 (dark blue; \textit{h}-BN close by) and 2 (light blue, no \textit{h}-BN close by) of the same nanowire with different lengths. The plasmon mode at around 180 meV disperses as a function of length and does not show peak splitting (the two peaks observed are the expected ones for \textit{h}-BN, superimposed on the nanowire dipolar plasmon peak), the behavior expected for the uncoupled system.}
\end{figure*}

\begin{figure*}
\includegraphics[width=9cm]{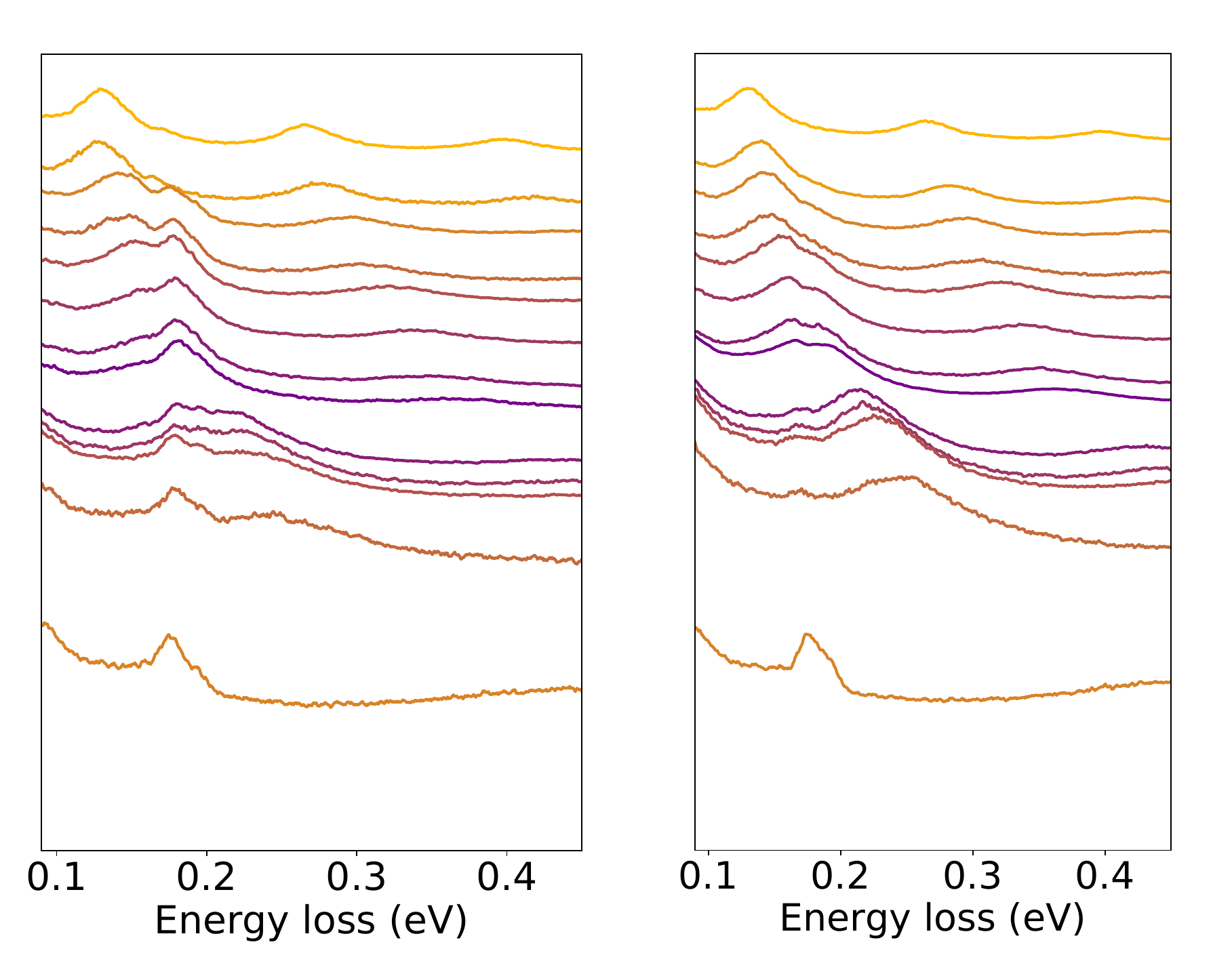}
\caption{\textbf{Nanowire with intermediary amount of \textit{h}-BN, 17 meV splitting}}
\end{figure*}

\begin{figure*}
\includegraphics[width=11cm]{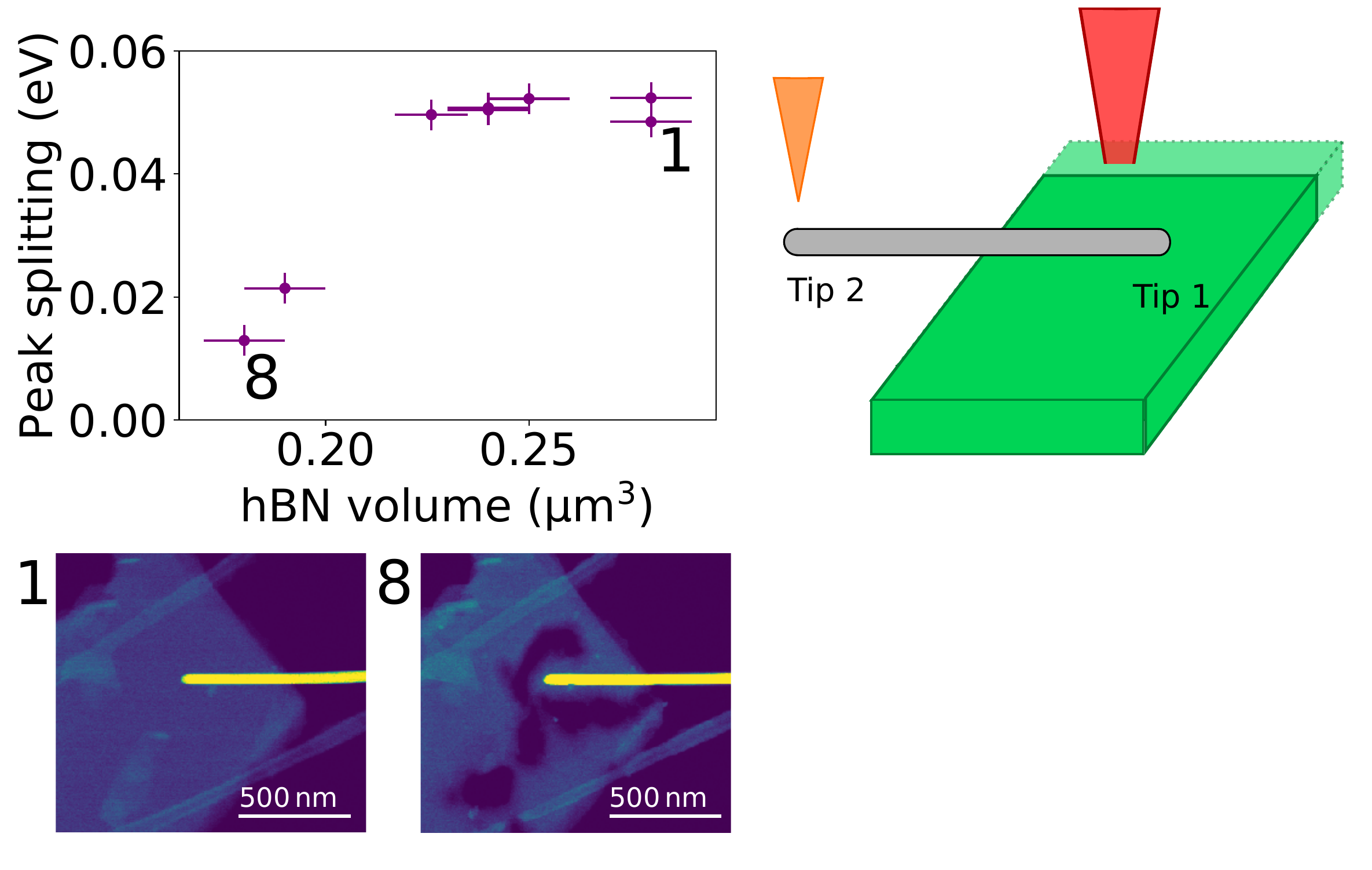}
\caption{\textbf{Coupling strength as a function of \textit{h}-BN volume close to nanowire tip.} The splitting of the two strong-coupled modes of a nanowire changes as a function of the \textit{h}-BN volume present on the other tip of nanowire. The volume was measured within 500 nm of the nanowire tip. For this specific nanowire the splitting changed from 57 meV to 12 meV. The change in the projection of the \textit{h}-BN volume between steps 1 and 8 is shown in the images below, marked “1” and “8”. Milling of \textit{h}-BN was achieved using a 200 keV high current ($\mu$A) beam in a conventional TEM.
}
\end{figure*}

\begin{figure*}
\includegraphics[width=9cm]{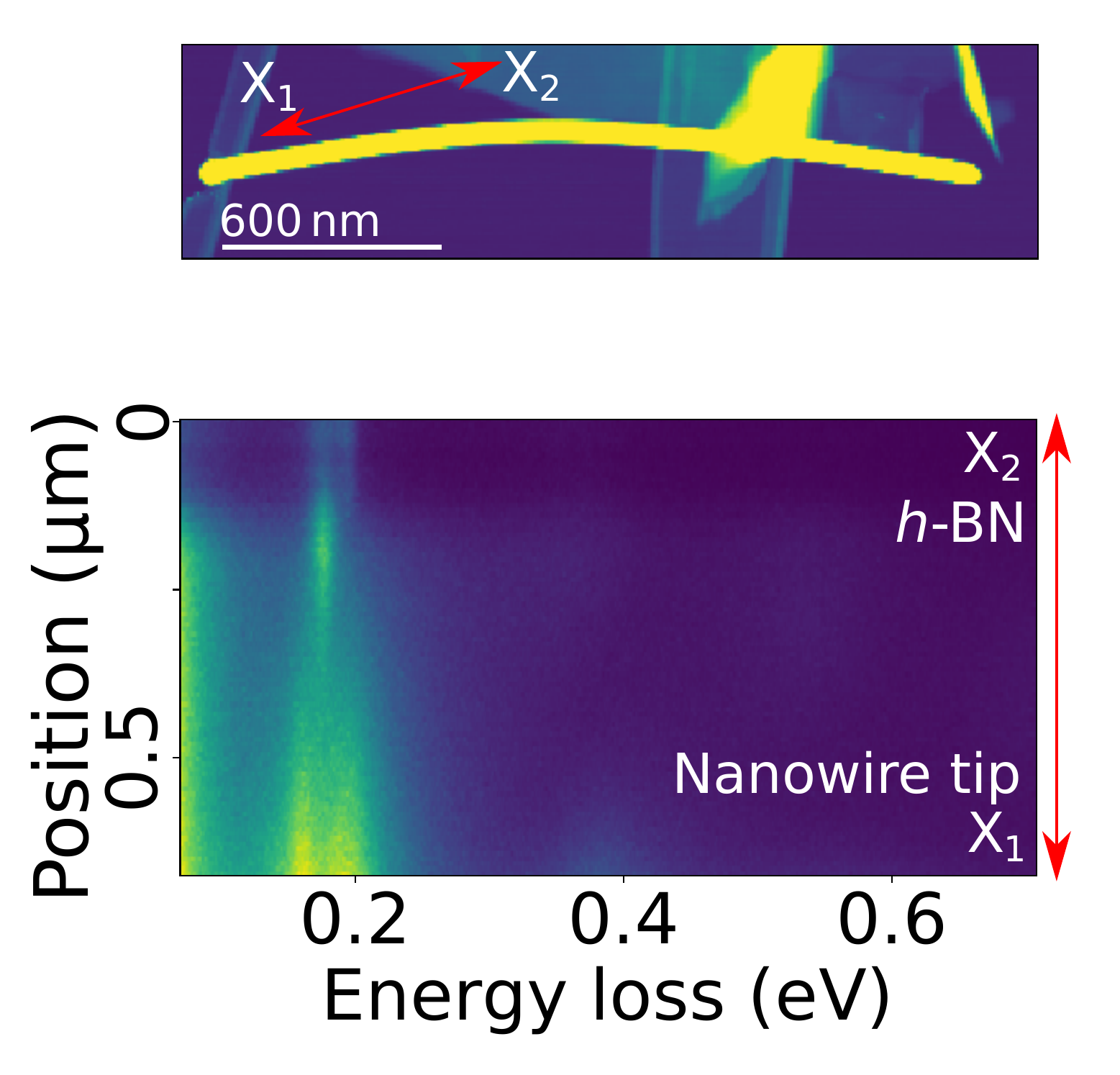}
\caption{\textbf{Spatial-spectral evolution around the tip of the nanowire when strong coupling occurs.}  Spatial evolution of the EELS intensity when moving the electron beam from the tip of the nanowire toward and into the \textit{h}-BN flake along the blue arrow marked X1-X2. The two strongly coupled modes lose intensity away from the nanowire tip as the aloof spectra of \textit{h}-BN appears. With the electron beam in \textit{h}-BN, the two Fuchs-Kliewer modes are observed.}
\end{figure*}

\begin{figure*}
\includegraphics[width=12cm]{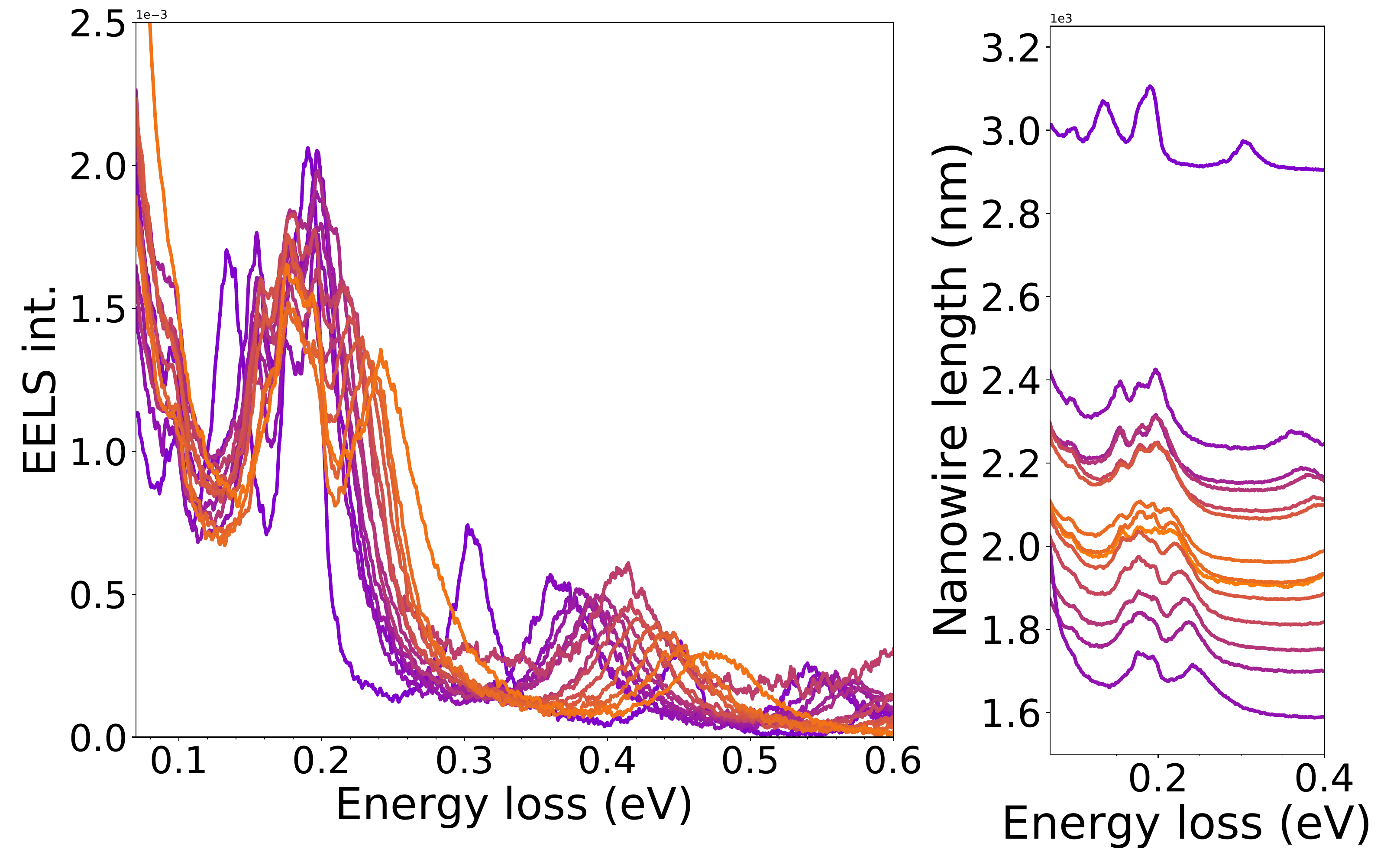}
\caption{\textbf{ Spectral evolution for all lengths of the nanowire on h-BN used in Fig. 3 of the main text.} The length of nanowire increases towards the top from 1.6 $\mu$m to 2.9 $\mu$m. The left panel shows all the spectra superimposed and the right panel offset based on the nanowire length.}
\end{figure*}

\begin{figure*}
\includegraphics[width=9cm]{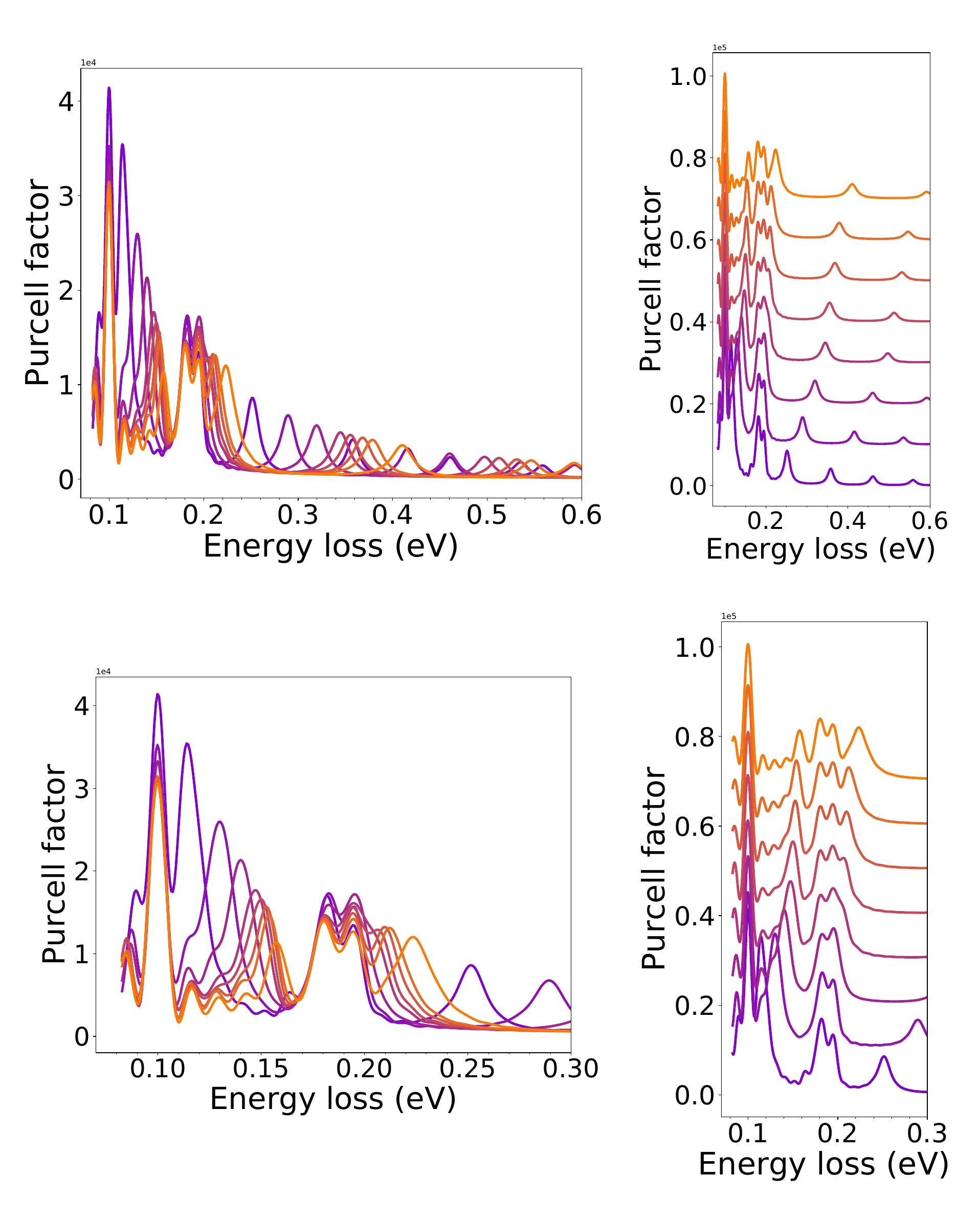}
\caption{\textbf{Numerical calculations of the Purcell factor due to a Ag nanowire placed on a 50 nm thick \textit{h}-BN layer.}}
\end{figure*}

\begin{figure*}
\includegraphics[width=9cm]{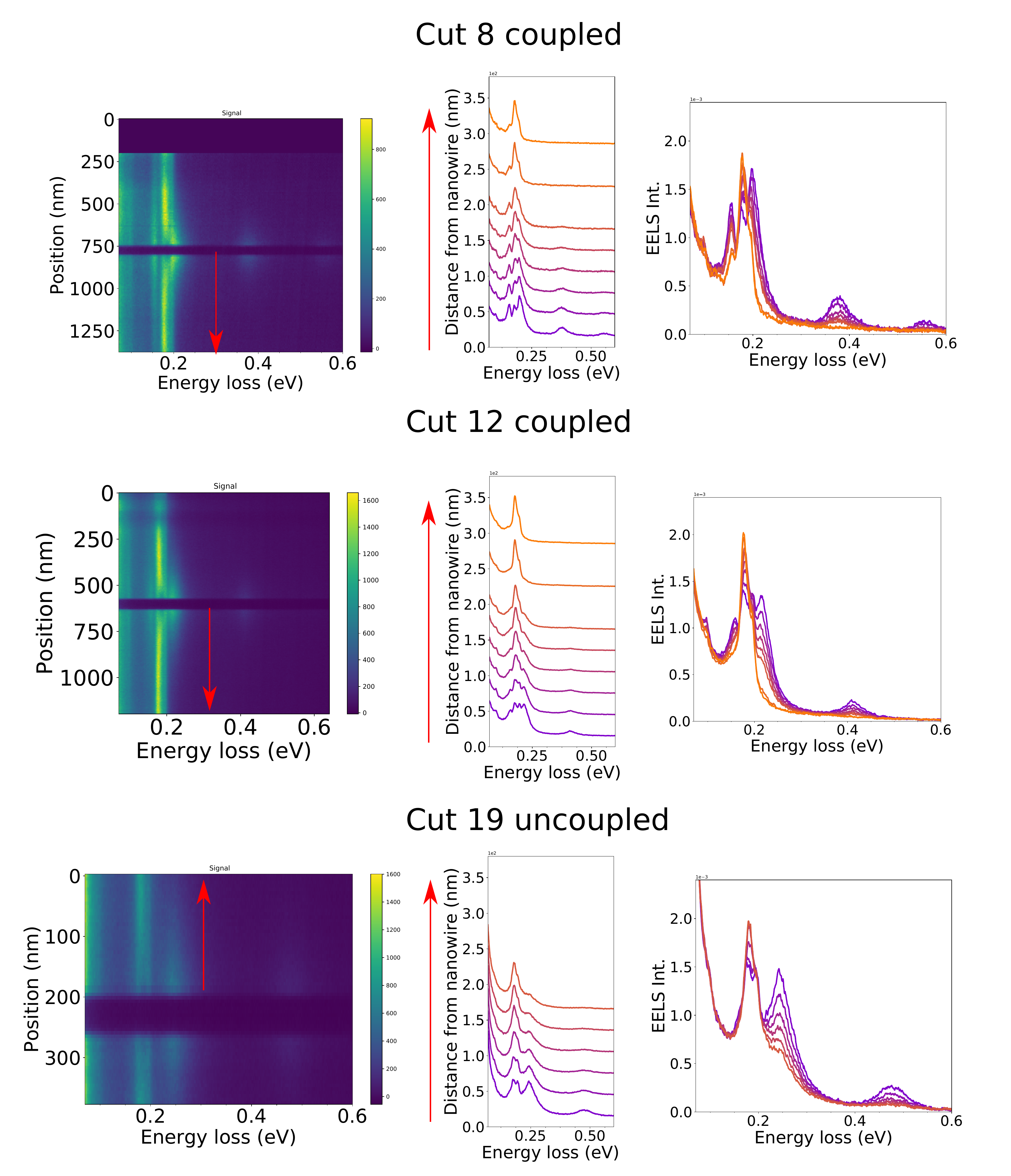}
\caption{\textbf{Spectral evolution away from the surface of the nanowire for different nanowire length.} The left panel shows a 2D map of EELS intensity as a function of energy loss and distance to the nanowire (the nanowire is positioned at the dark bands in the center of each panel. The center and left panel shows the spectral evolution as a function of distance for selected distances with and without an offset (given by the distance), respectively.}
\end{figure*}

\begin{figure*}
\includegraphics[width=9cm]{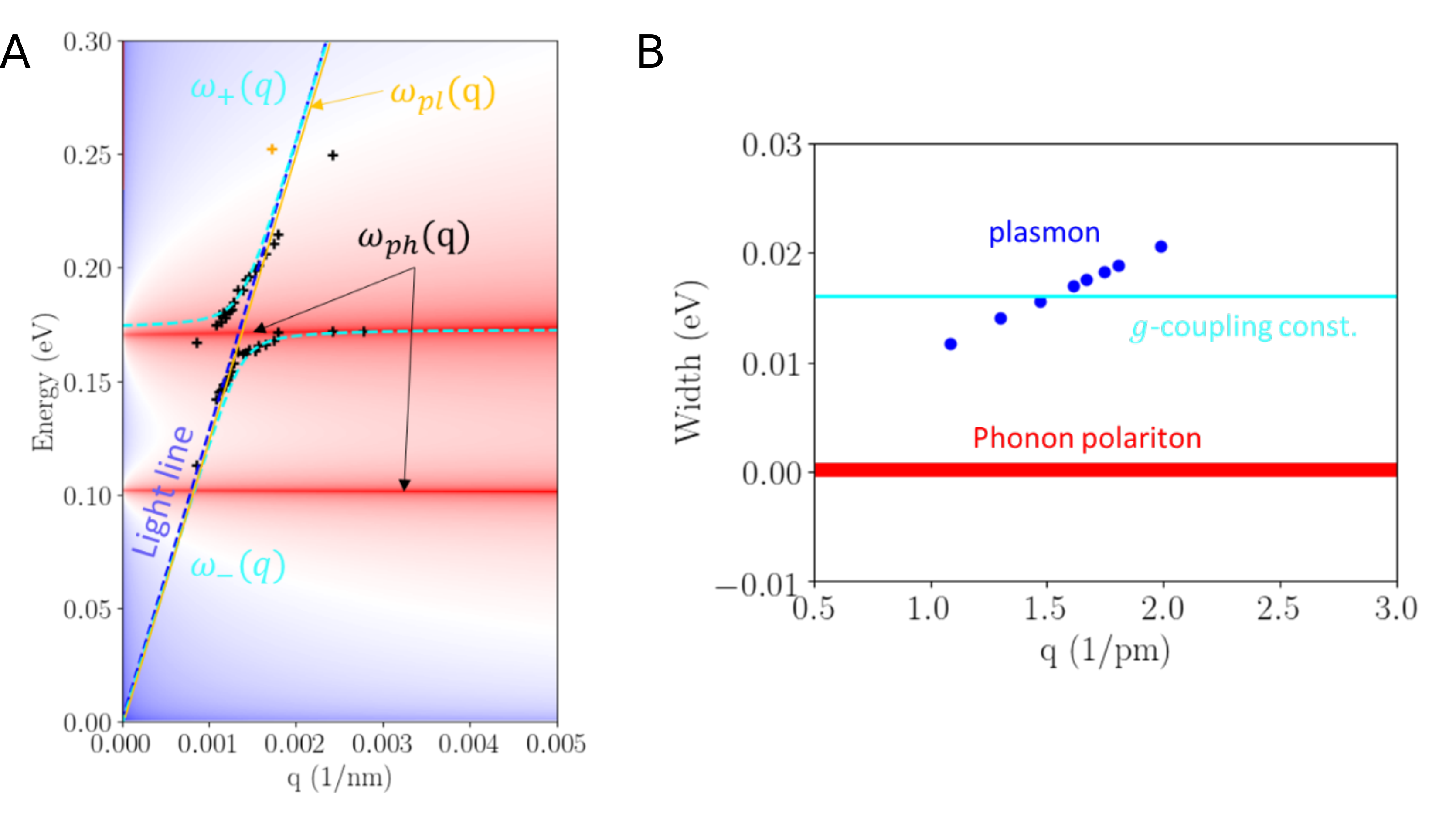}
\caption{\textbf{(A)} Strong coupling calculation between plasmon and phonon modes as described in the methods above. \textbf{(B)} Comparison between the coupling strength $g$ to the plasmon and phonon widths extracted from simulations.}
\end{figure*}

\end{document}